\pdfoutput=1 
\documentclass{JINST}
\let\ifpdf\relax 
\usepackage{txfonts}

\title{The Micromegas Project for the ATLAS Upgrade}

\author{G. Iakovidis$^{a,b}$\thanks{On behalf of the MAMMA collaboration.}~\\
\llap{$^a$}National Technical University of Athens\\
  Zografou Campus, GR15773, Athens, Greece\\
\llap{$^b$}Brookhaven National Laboratory\\
  Upton, NY 11973, USA\\
E-mail: \email{george.iakovidis@cern.ch}}

\abstract{Micromegas is one of the detector technologies (along with the small Thin Gap Chambers) that has been chosen for precision tracking and triggering purposes of the ATLAS muon forward detectors in the view of LHC luminosity increase. To fulfill the requirements of such upgrade, several prototype micromegas detectors were tested in recent test beam campaigns with high energy hadron beams at CERN. Performance studies and results on spatial resolution for perpendicular and inclined tracks, efficiency, as well detector performance and comparison to simulation in a magnetic field are presented .Moreover, an overview of detector performance after neutron, X-ray, gammas and alphas exposure and construction achievements of large area micromegas detectors are presented.}


\begin{document}
\sloppy

\section{Introduction}\label{sec:intro}
The Large Hadron Collider (LHC) at CERN, after the scheduled shutdown of 2017-2018, will resume its operation with a luminosity increase of five times its original design luminosity of $\mathcal{L}=10^{34}\,\mathrm{cm^{-2}s^{-1}}$. For the ATLAS detector (ref. \cite{cite:atlastdr}), such a luminosity increase means higher particle rates. While in most of the ATLAS muon system the detectors have enough safety margin to handle these rates, the first forward station of the muon spectrometer, called the Small Wheel, will exceed its design capabilities. At pseudorapidity $\eta = \pm2.7$, rates up to $15\,\mathrm{kHz/cm^{2}}$ are expected, far higher than what the currently installed detectors can handle. Furthermore, the upgraded Small Wheel is expected to take part in the Level 1 trigger decision, something that the present system was not designed for. The physics objective is to sharpen the trigger threshold turn-on as well as discriminate against background while maintaining the low transverse momentum ($p_{\mathrm{T}}$) threshold for single leptons ($e$ and $\mu$) and keeping the Level-1 rate at a manageable level.

The  Muon ATLAS MicroMegas Activity (MAMMA) R\&D explored the potential of the micromegas technology for its use in LHC detectors and finally proposed to equip the New Small Wheel (NSW) with micromegas detectors, combining trigger and precision tracking functionality in a single device. In middle 2013 the ATLAS Collaboration endorsed the proposal (ref. \cite{cite:tdr}). In total, eight planes of micromegas (ref. \cite{cite:Giomataris}) detectors covering the full NSW will be installed, corresponding to a total detector area of 1200\,$\mathrm{m^2}$. In addition to the micromegas, the NSW should also be equipped with eight planes of thin-gap multiwire detectors, called sTGC, such as to create a fully redundant system, both for trigger and tracking.

\section{Micromegas Detectors}\label{sec:largeMM}
The first two years (2008/2009) of the MAMMA activities were focused on building several micromegas detectors using the relatively new bulk technique (ref. \cite{cite:Giomataris}) at the CERN PCB workshop. Different layouts, with respect to strip pitch and strip length and gas mixtures were tested. Detectors with strip length up to 1\,m were build while the operating gas mixture was settled on a $\mathrm{Ar:CO_2}$ 93:7 mixture (used widely among other muon detectors in ATLAS).

\begin{figure}[tbp] 
\centering
\includegraphics[width=1.\textwidth]{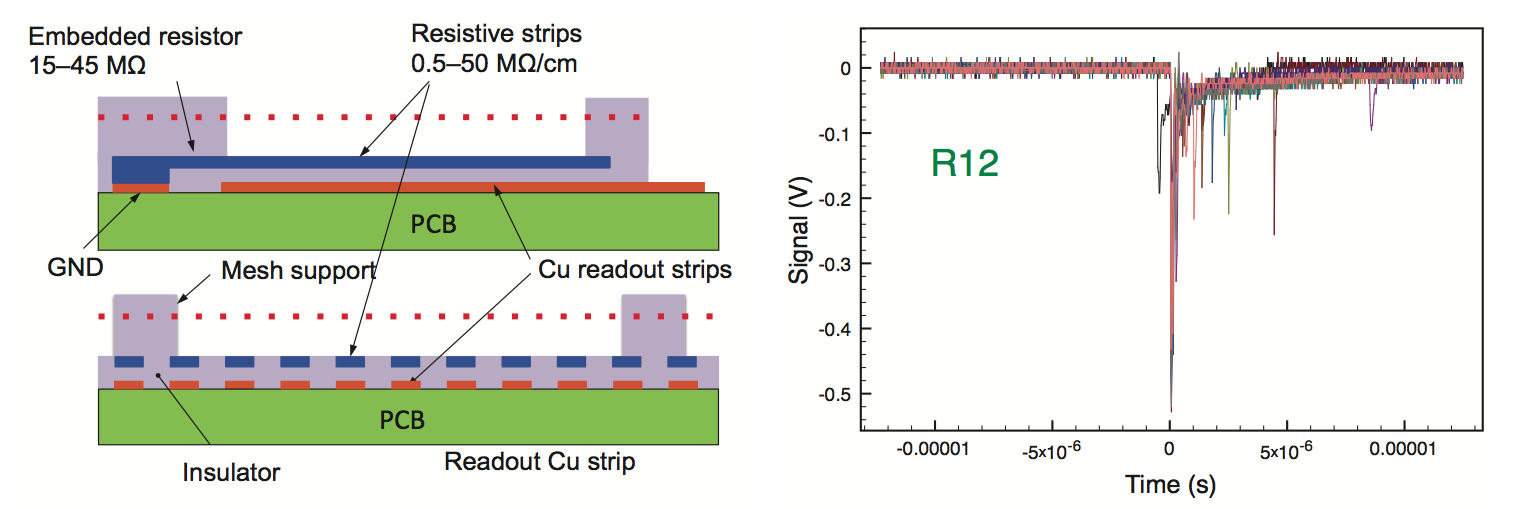}
\caption{On the left, a sketch (not in scale) of the resistive-strip protection principle with a view along and orthogonal to the strip direction; on the right, oscilloscope screen shot of about 10 spark signals as seen on the readout strips with 50\,$\mathrm{\Omega}$ termination.}
\label{fig:resistiveMM}
\end{figure}

\begin{figure}[tbp] 
\centering
\includegraphics[width=1.\textwidth]{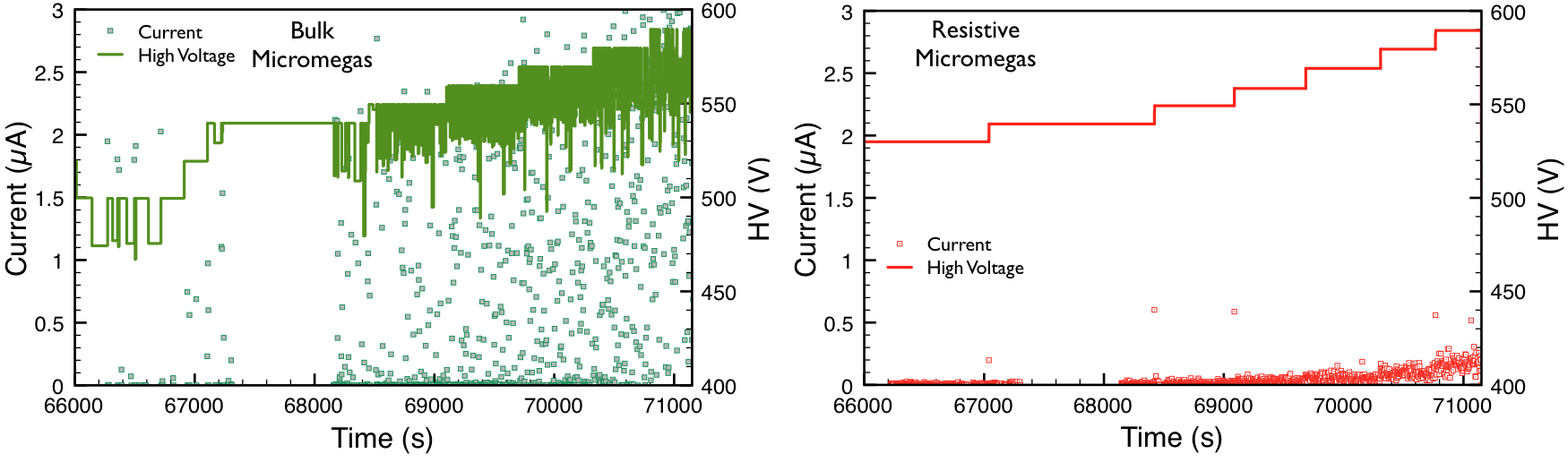}
\caption{Monitored HV (continuous line) and current (points) as a function of HV mesh under neutron irradiation, left a non-resistive micromegas; right a micromegas with resistive-strip protection layer.}
\label{fig:micromegasPerformance}
\end{figure}

The main issue for using micromegas detectors in high energy experiment like ATLAS addressed to be sparks. Sparks lead to HV breakdowns with relatively long recovery times and sometimes damage to the detector itself. Most of the MAMMA R\&D activities since 2010, were dedicated to address the spark issue. Many different approaches were tried. The final approach was settled on a scheme with a layer of resistive strips above the readout strips that matches geometrically them. This technique transformed micromegas detectors spark resistant while maintaining their ability to measure minimum-ionizing particles with excellent precision in high-rate environments. Resistive strips layout rather than a continuous resistive layer was chosen mainly to avoid charge spreading across several readout strips while keeping the area affected by a discharge as small as possible and thus maintaining a high rate capability of the detector. The principle of the resistive-strip protection scheme is illustrated in figure \ref{fig:resistiveMM} and described in more detail in ref. \cite{cite:AlexMM}. The power of the resistive-strip spark protection scheme is illustrated in figure \ref{fig:micromegasPerformance}. It shows the monitored HV and the currents for a standard micromegas and one with the resistive-strip protection under neutron irradiation (ref. \cite{cite:AlexNeutrons}\cite{cite:Iakovidis}) for different mesh HV settings.

\subsection{ATLAS Small Wheel Upgrade}\label{sec:wheels}
\begin{figure}[tbp] 
\centering
\includegraphics[width=.38\textwidth]{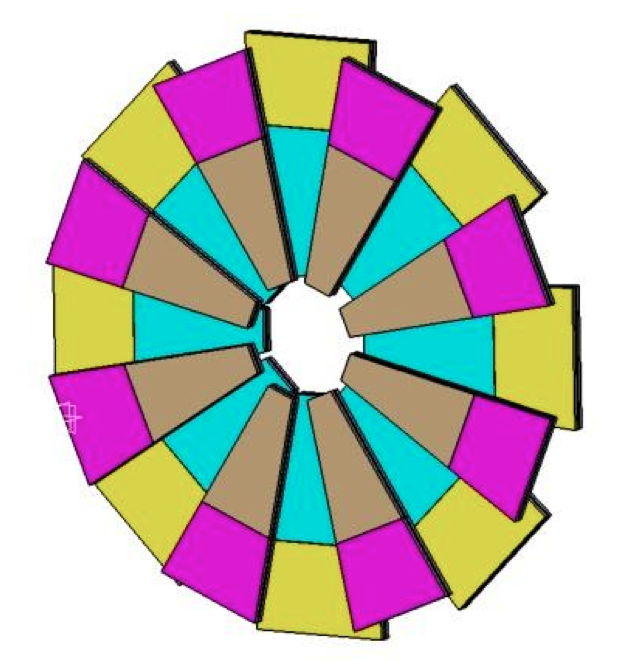}
\includegraphics[width=.58\textwidth]{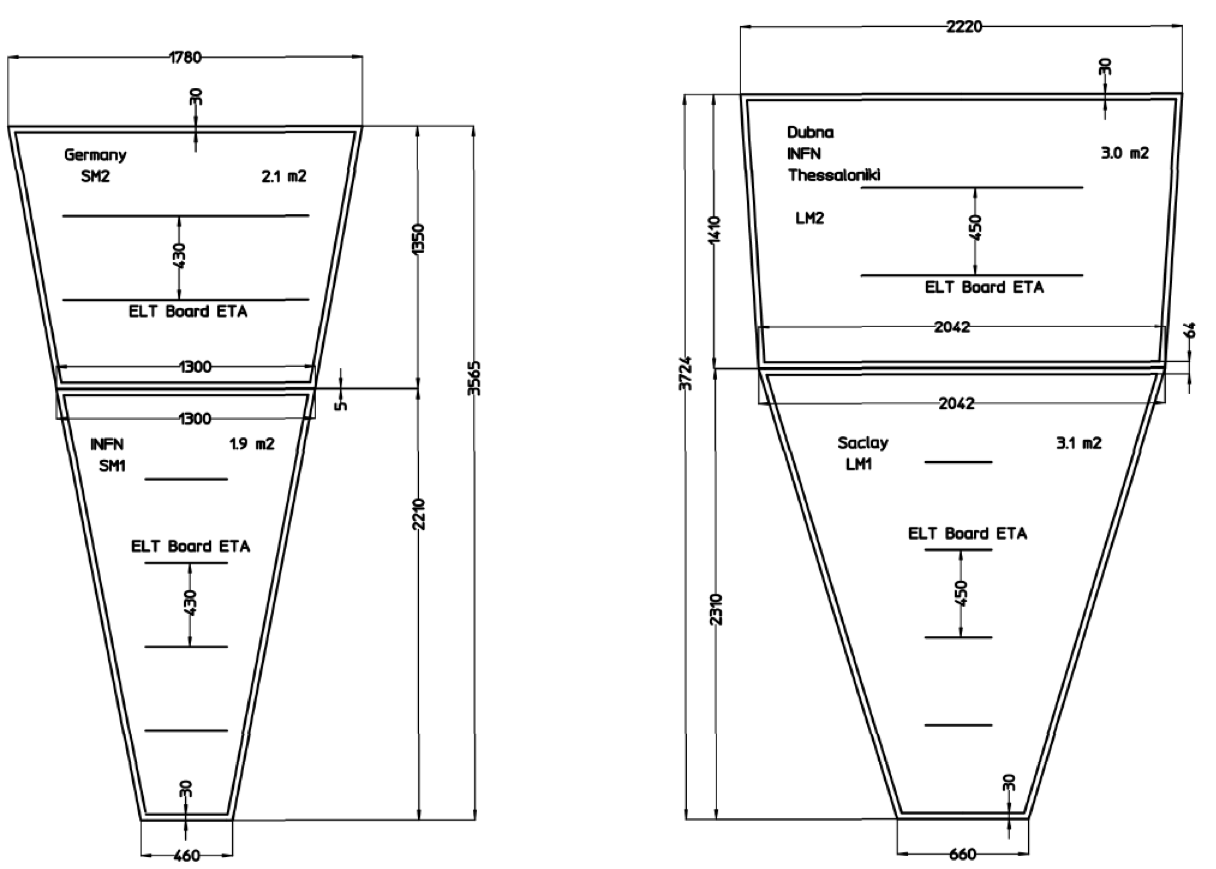}

\caption{On the left the NSW layout is shown. On the middle and right the segmentation of the modules of small and large sector respectively. }
\label{fig:wheels}
\end{figure}

Figure \ref{fig:wheels} shows the NSW (ref. \cite{cite:tdr}) layout divided into small and large sectors. Every sector consists eight layers of PCB boards each one divided into two micromegas. In total 512 micromegas up to a surface of 3.1\,m$^2$ should be build covering an area of 1200\,$\mathrm{m^2}$. The strip pitch of $\sim$450\,$\muup$m results in a system of a total of 2.1\,M channels. Strips on the four out of the eight layers will be under an angle of $\pm1.5^{\circ}$ providing second coordinate measurement while they contribute to the precision coordinate measurement with the other four at the same time. 

\subsection{Large-Area Micromegas}\label{sec:large}
Over the last few years several micromegas prototypes of an active area of $\sim0.5\,\mathrm{m^2}$ were build (ref. \cite{cite:AlexLarge}\cite{cite:joergLarge}) using the bulk technique. Experience with these detectors shows that mechanically the mesh deposition process is well under control resulting in a very homogeneous detector response over the full area. The weak point lies in the risk of enclosing some impurities (dust or dirt) under the mesh, leading to a high-ohmic current bridge between the mesh and the resistive strips. Cleaning out such dust particles is difficult or impossible. Overcoming this issue, a novel construction technique was developed over the last year. The chambers were constructed with a removable mesh. It consists of the readout panel and the drift electrode panel, the latter incorporates the mesh. Details on this construction method can be found in ref. \cite{cite:joergLarge}. 

\begin{figure}[tbp] 
\centering
\includegraphics[width=.47\textwidth]{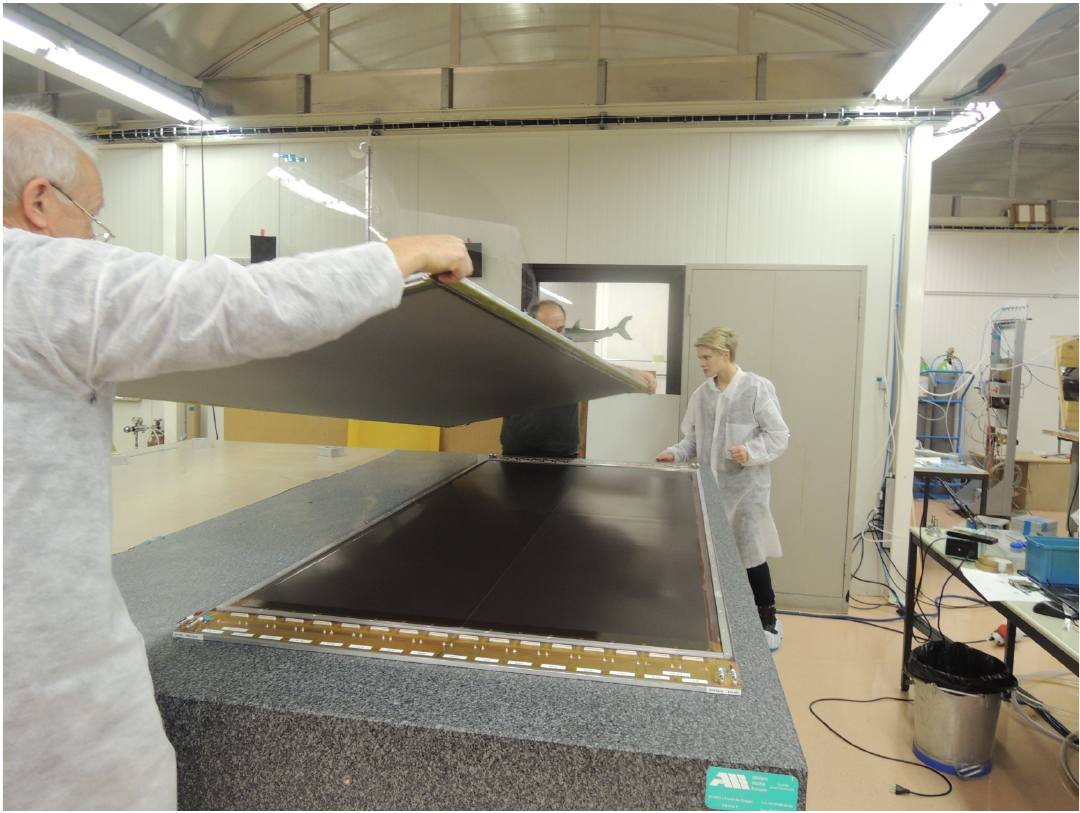}
\includegraphics[width=.47\textwidth]{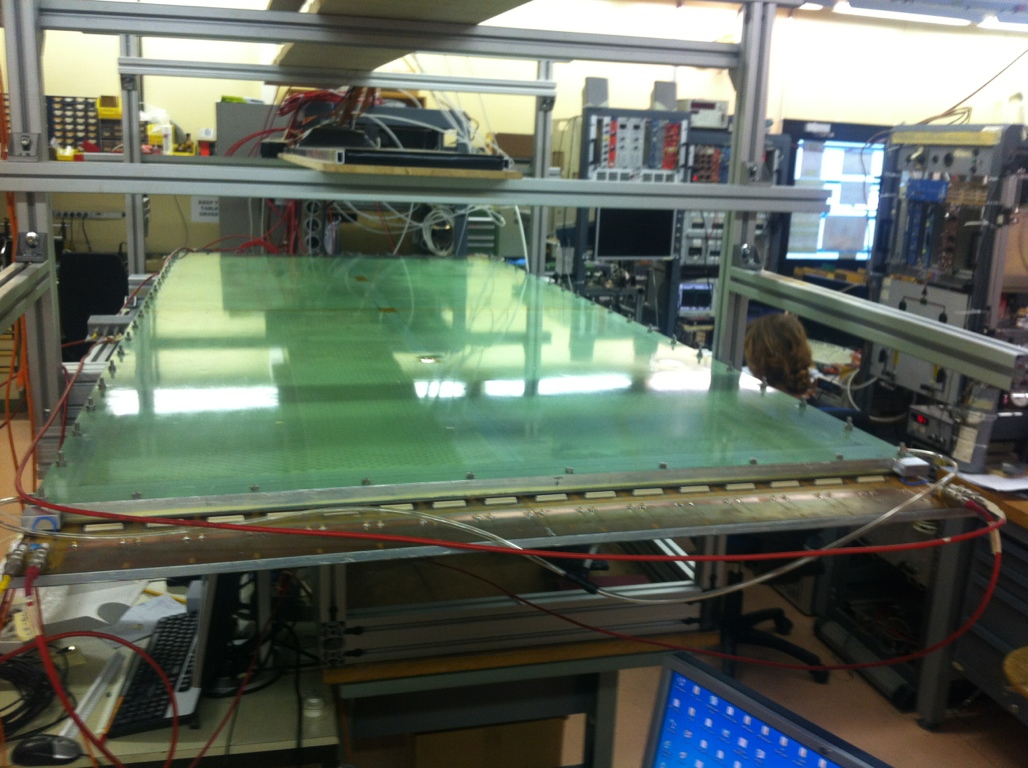}

\caption{On the left, the assembly of the large $1\times2.4\,\mathrm{m^2}$ micromegas prototype. On the right, the large prototype sitting on the cosmic stand at the CERN RD51 lab.}
\label{fig:micromegasPrototype}
\end{figure}

\begin{figure}[tbp] 
\centering
\includegraphics[width=.47\textwidth]{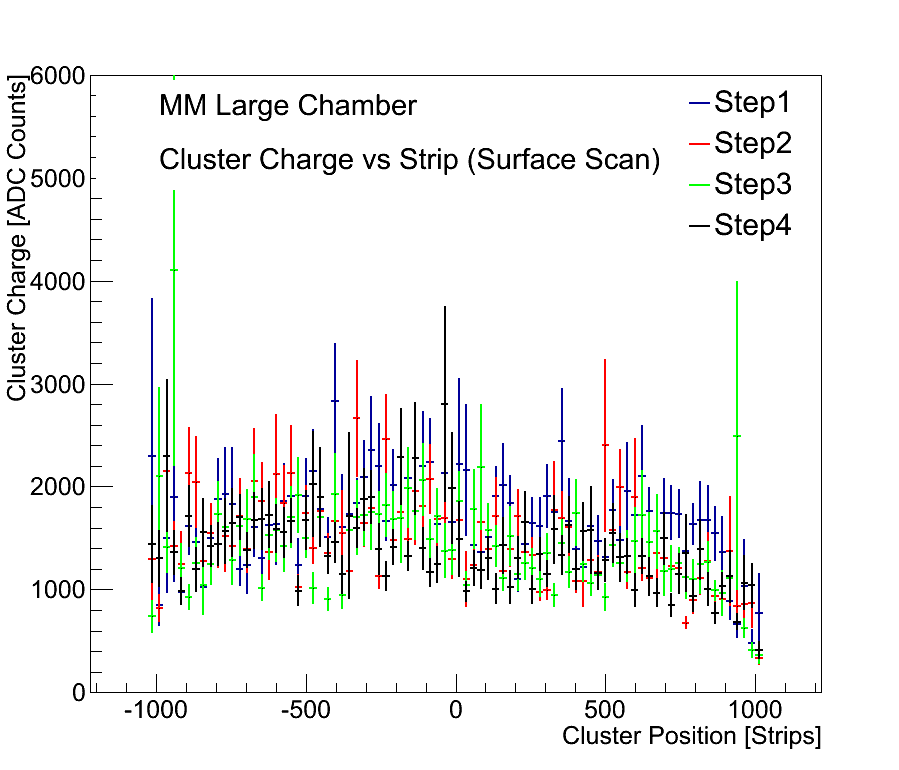}
\includegraphics[width=.47\textwidth]{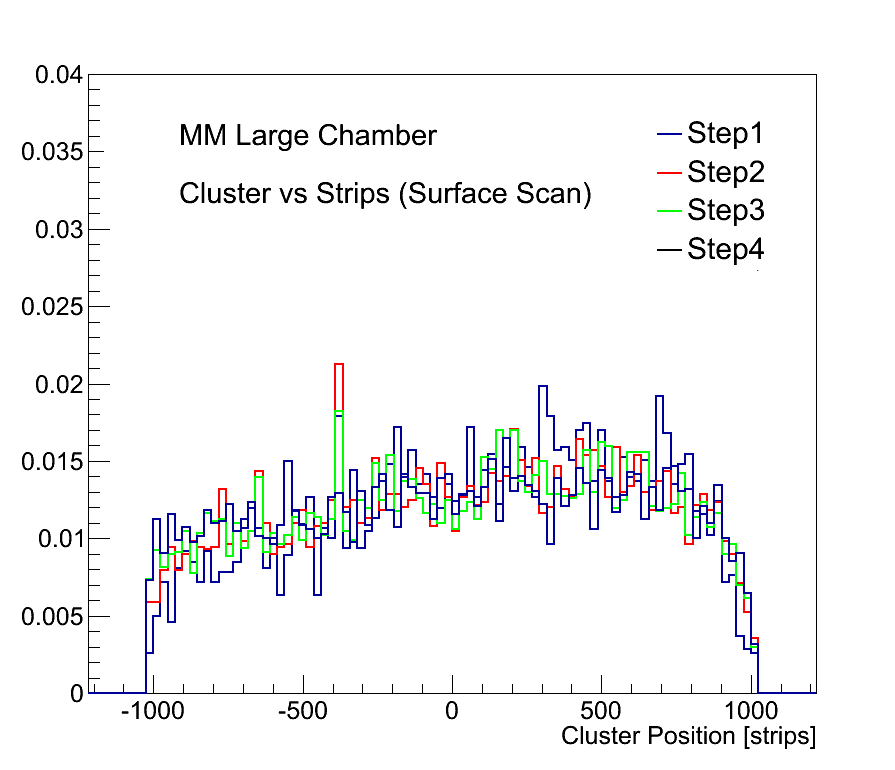}

\caption{The reconstructed cluster charge (left) and position (right) over the full detector surface.}
\label{fig:micromegasPrototypePerf}
\end{figure}

In the beginning of 2013 two large $1\times2.4\,\mathrm{m^2}$ ($0.92\times2.12\,\mathrm{m^2}$ active area) micromegas detectors were build. The readout panel is composed by four micromegas boards of 0.5\,mm thick glued together to an aluminum plate that served as a stiffening panel. The total number of strips are $2\times2048$. On top of the resistive strips 128 $\muup$m high pillars were deposited every 4\,mm, following the same procedure as used in the bulk process. On the drift panel the ``inner'' skin serves as drift electrode while on the same skin, a 5\,mm high frame is fixed to which the mesh is glued. When the chamber is closed the mesh, that makes the ground contact, sits on the 128\,$\muup$m high pillars. When HV is applied to the resistive strips the electrostatic force between mesh and resistive strips ensures that the mesh is in good contact with the pillars. Data taken with the chamber in the test beam show an excellent performance with clean signals and a homogeneous response over the full area of the detector. Figure \ref{fig:micromegasPrototype} shows the detector on the assembly and testing phases while the figure \ref{fig:micromegasPrototypePerf} shows the homogeneity over its surface.

\section{Performance}\label{sec:performanceMM}
Micromegas detectors of an active surface of 10$\times$10\,cm$^2$ have been tested during test beam campaigns at CERN with high momentum hadron beams. Using those detectors with a strip pitch of 400\,$\muup$m a spatial resolution of 65\,$\muup$m was easily achieved for perpendicular tracks. Since the NSW will be located in the ATLAS experiment tracks between 10$^{\circ}$--30$^{\circ}$ are expected, studying the performance with inclined tracks is of particular interest.

\subsection{The $\muup$TPC Scheme}\label{sec:resolution}
\begin{figure}[tbp] 
\centering
  $\vcenter{\hbox{\includegraphics[width=.4\textwidth]{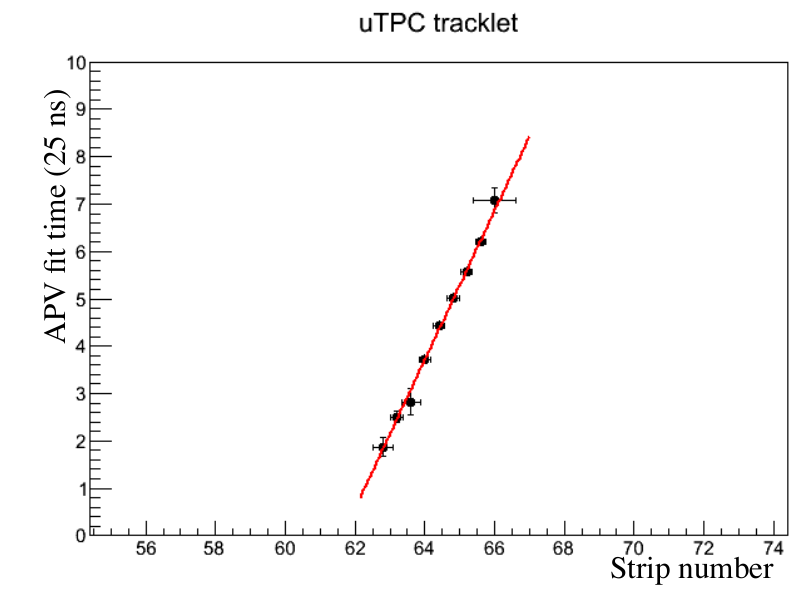}}}$
  \hspace*{.2in}
  $\vcenter{\hbox{\includegraphics[width=.45\textwidth]{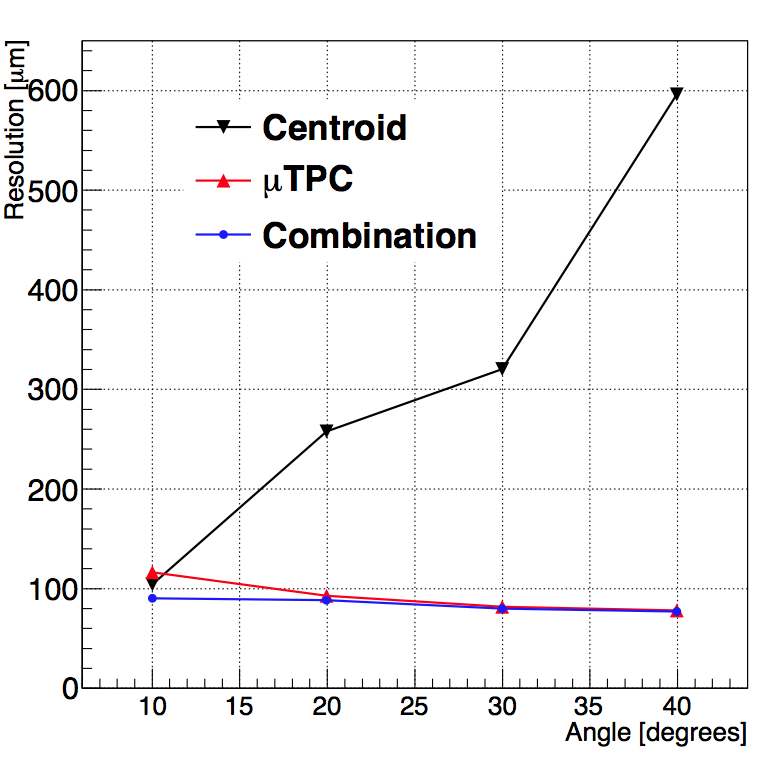}}}$
\caption{On the left a reconstructed track in the 5\,mm drift gap under 30$^{\circ}$. On the right plot the spatial resolution versus the incident track angle with different reconstruction techniques.}
\label{fig:uTPC}
\end{figure}

Measuring the arrival time of the ionized electrons with a time resolution of a few nanoseconds allows reconstructing the position of the ionization process and thus reconstruction of the particle track in the drift gap of the detector. With the Ar:CO$_2$ (93:7) mixture and an electrical drift field of 600\,V/cm the drift velocity is 4.7\,cm/$\muup$s, corresponding to a maximum drift time of about 100\,ns for a 5\,mm drift gap. Figure \ref{fig:uTPC} shows an example for a reconstructed track traversing the detector under $30^{\circ}$ and the spatial resolution as a function of the incident track angle using the $\muup$TPC mode. An analysis technique that combines the reconstructed $\muup$TPC and charge centroid points, improves the spatial resolution especially for particle tracks of around 10$^{\circ}$, while results in a homogeneous spatial resolution under 100\,$\muup$m along different track angles.

\subsection{Micromegas Inside Magnetic Field}\label{sec:bfield}
\begin{figure}[tbp]
\centering
\includegraphics[scale=0.044]{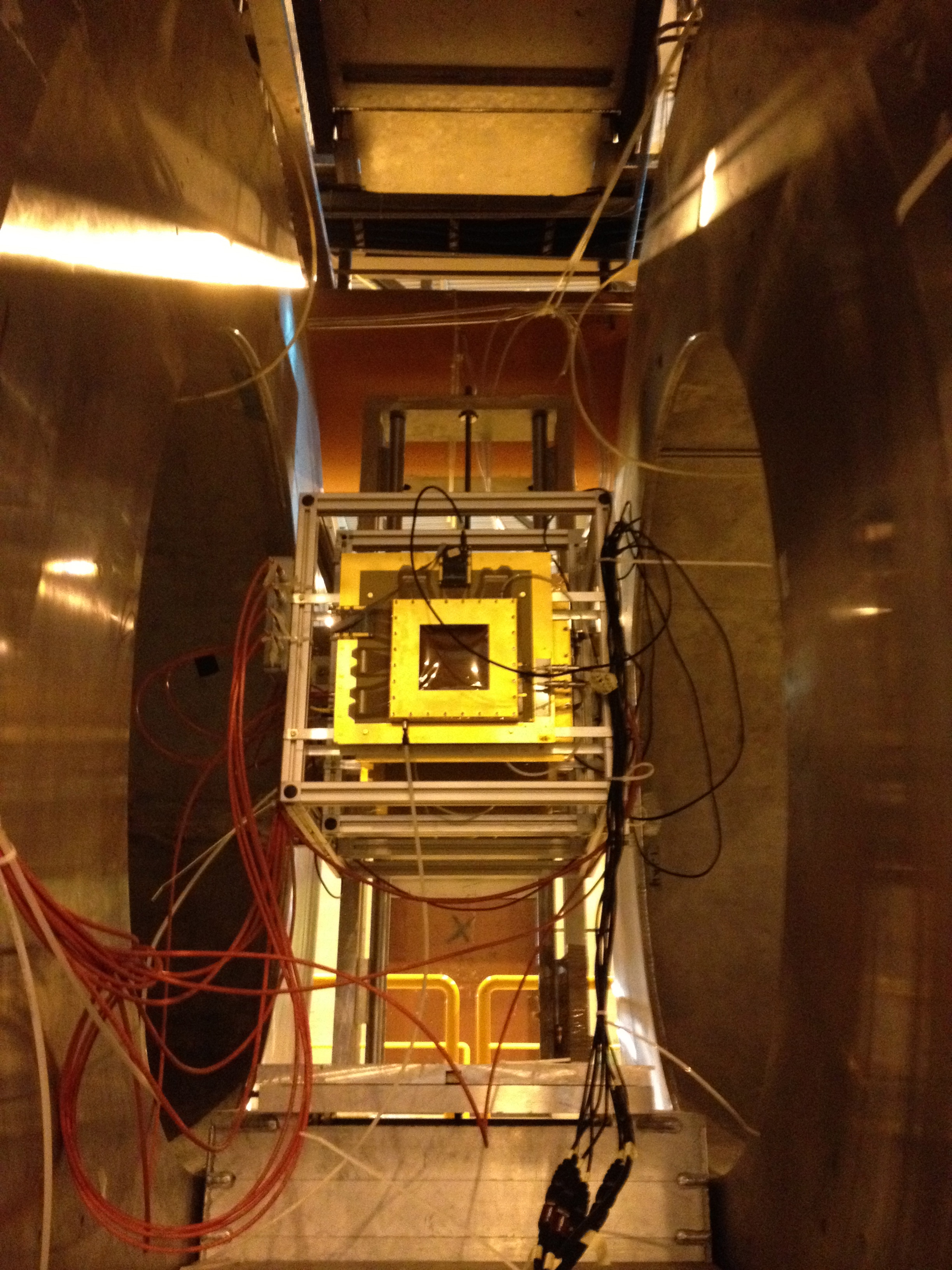}\hspace{20pt}  
\includegraphics[scale=0.0585]{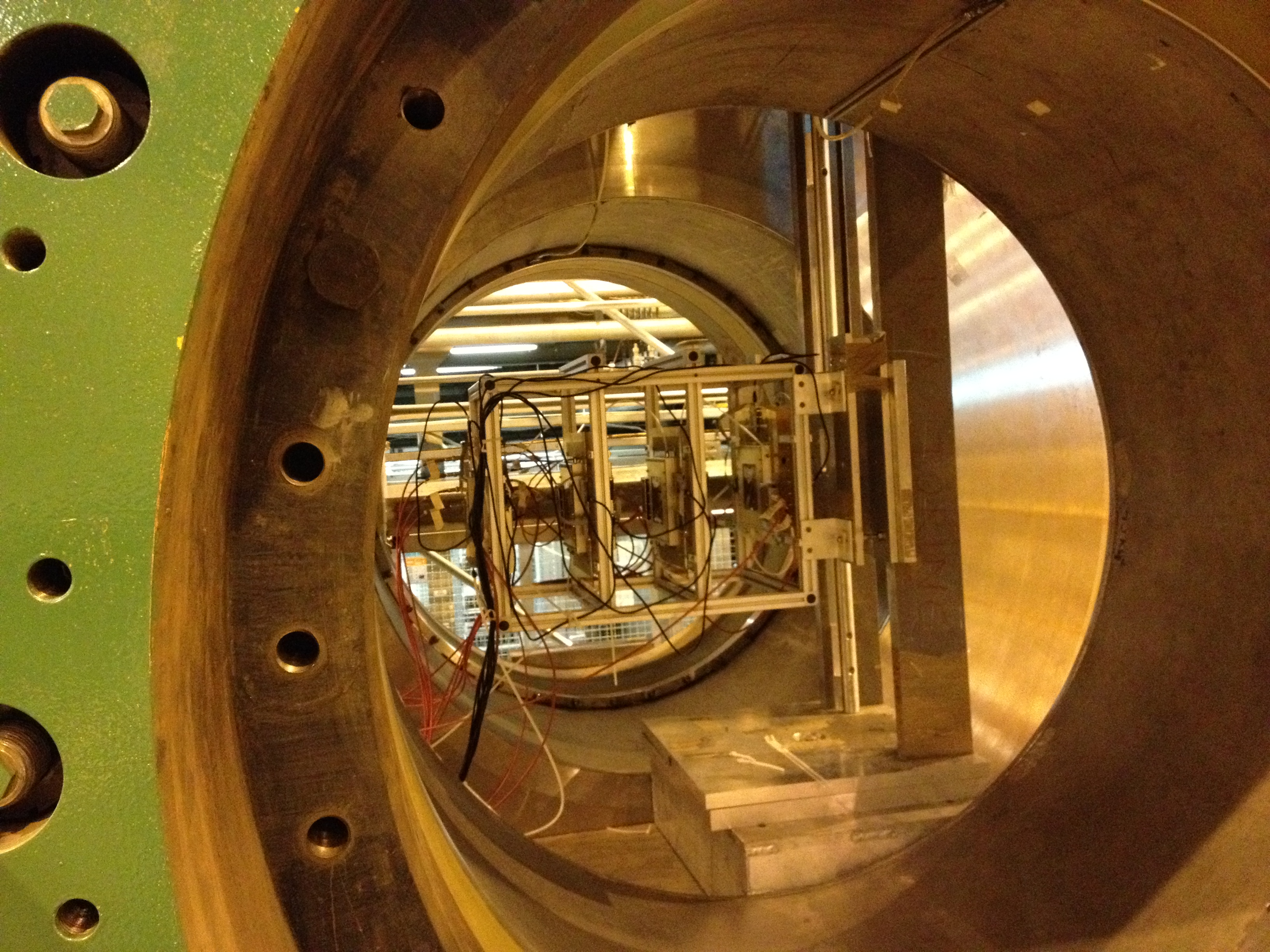}  
\caption{Eight migromegas chambers inside the superconducting magnet in H2 CERN on June 2012 test beam.} 
\label{fig_h2_photo}
\end{figure}

\begin{figure}[tbp] 
\centering
\includegraphics[width=.8\textwidth]{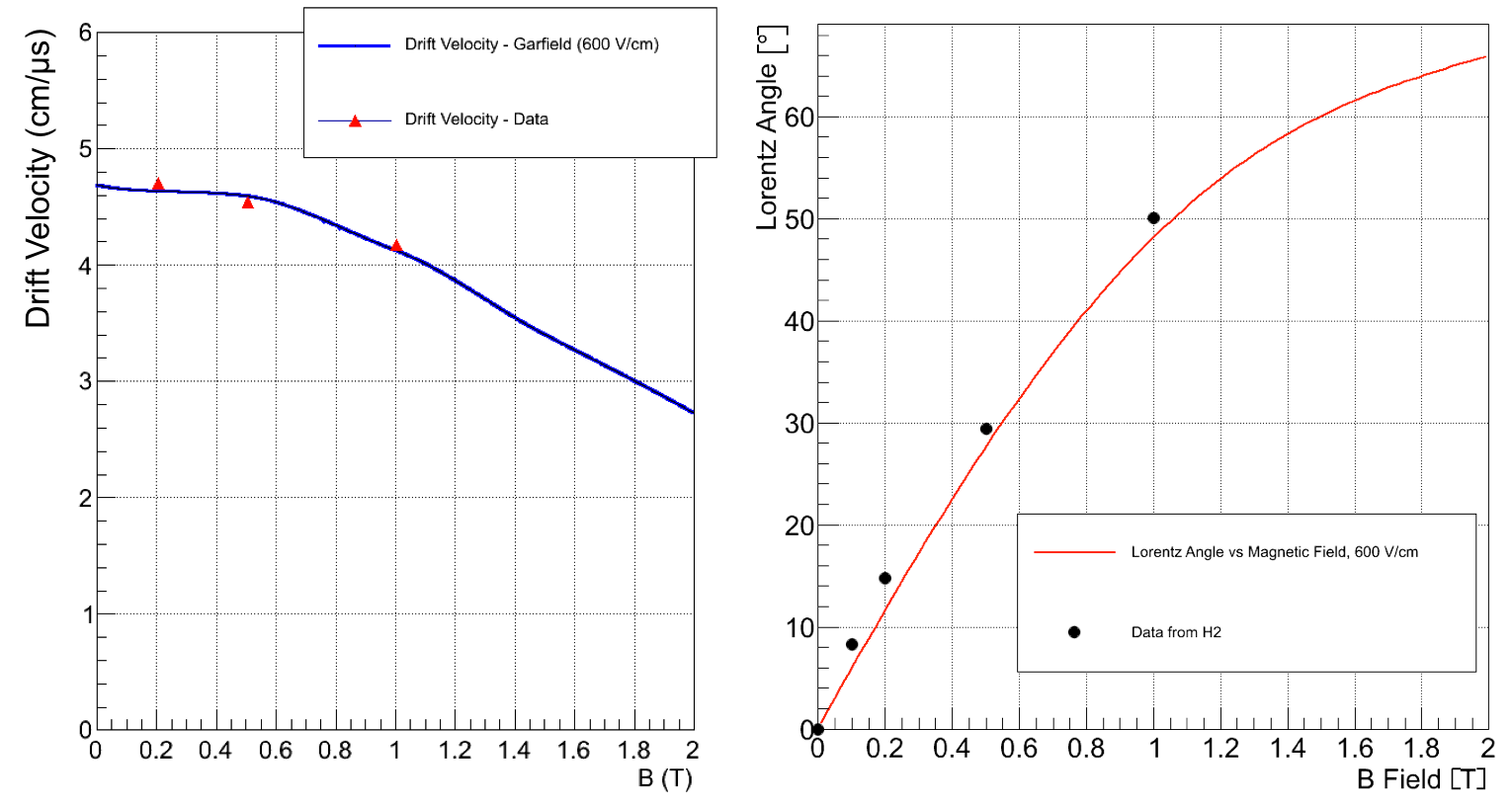}
\caption{The measured drift velocity (left plot, red points) and the lorentz angle measured (right plot, black points) versus the magnetic field. Both measurements were found to be in agreement with Garfield (ref. \cite{ref:garfield}) simulation.}
\label{fig:bfieldlorentz}
\end{figure}

The NSW will operate under a multi-directional and non constant magnetic field up to 0.5\,T. Several micromegas detectors of $10\times 10\,\mathrm{cm}^2$ active were installed inside a superconducting magnet at CERN (figure \ref{fig_h2_photo}) and tested under the influence of a magnetic field up to 2\,T. No degradation of the performance was observed while we were able to measure the Lorentz angle and the drift velocity. Figure \ref{fig:bfieldlorentz} shows those measurements in comparison with Monte Carlo simulation.  

\subsection{Triggering with micromegas}\label{sec:trigger}
Micromegas detector system will contribute to the formation of the Level-1 muon end-cap trigger, forming a very powerful, redundant trigger system along with sTGC detectors. Exploiting this capability,
however, requires a large number of electronics channels, about two million for an eight layer detector system. The electronics of the NSW must provide both a high resolution vector, in real time, to be used in
the formation of the muon Level-1 trigger in addition to the amplitude and time information provided on the
reception of a Level-1 accept.

\begin{figure}[tbp] 
\centering
\includegraphics[width=.5\textwidth]{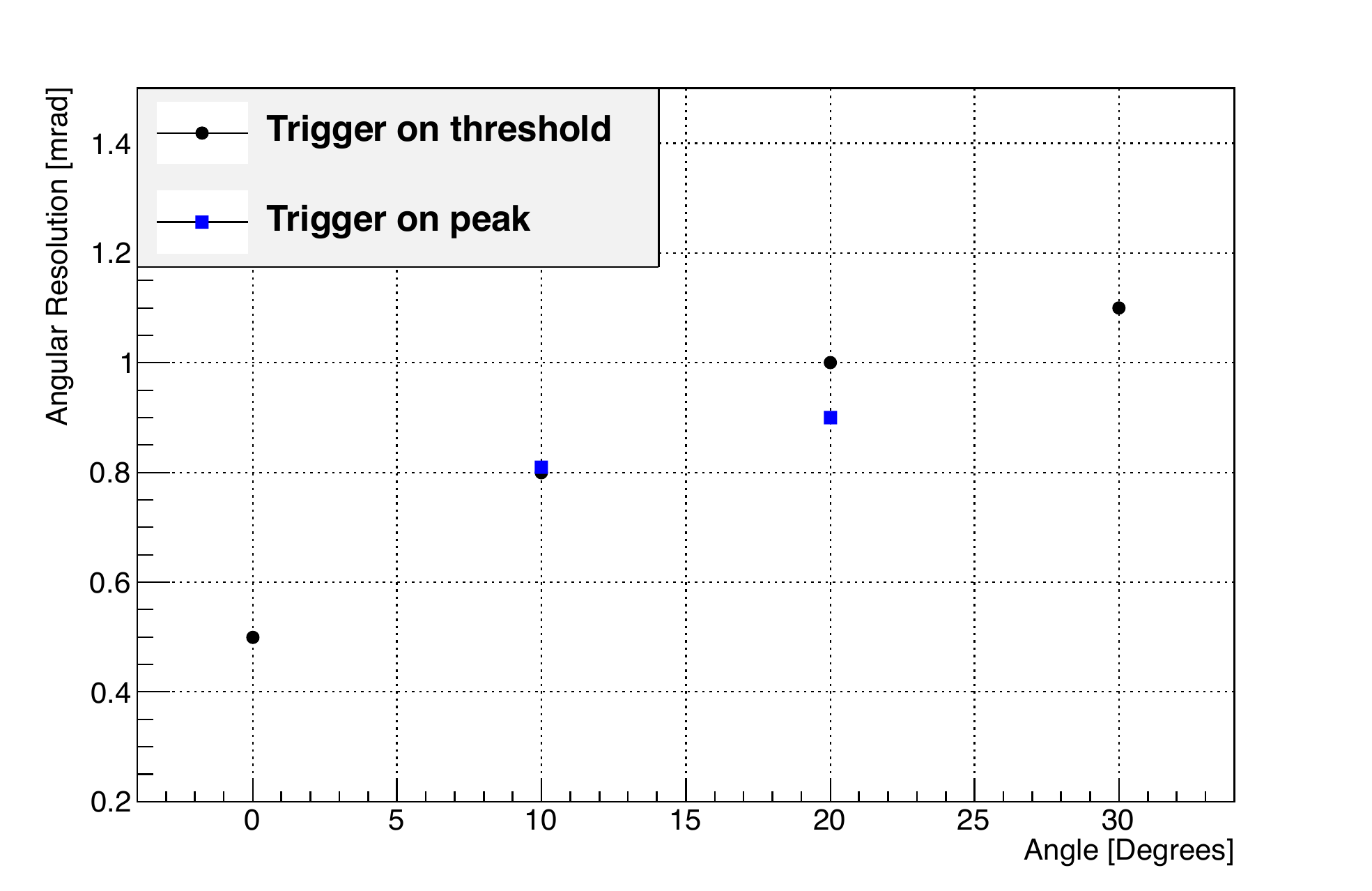}
\includegraphics[width=.4\textwidth]{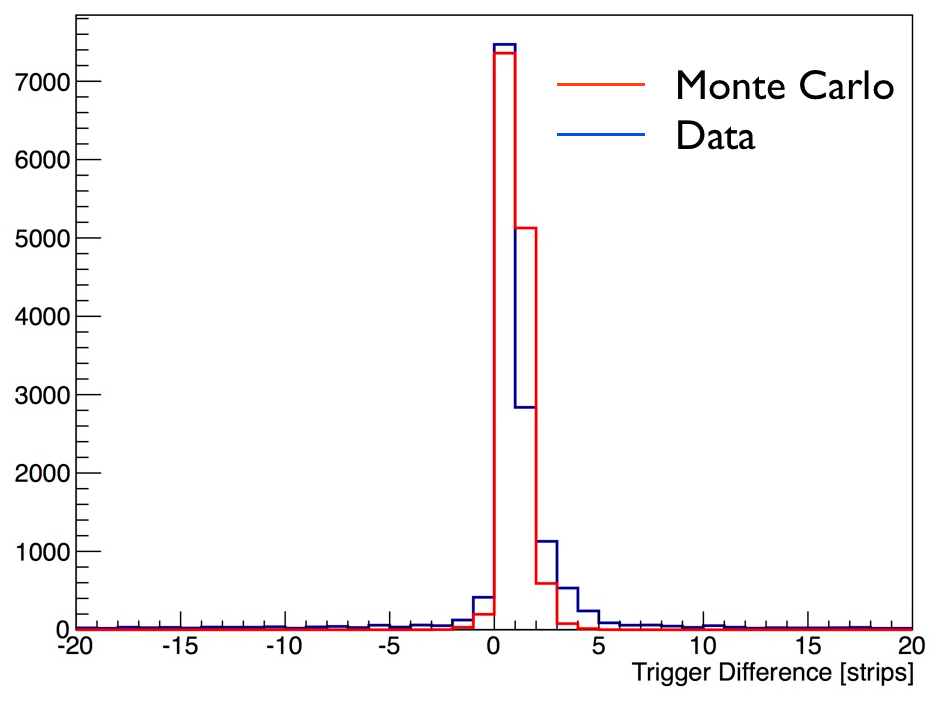}
\caption{On the left plot, the angular resolution with a set of 6 micromegas chambers with a level arm of 50\,cm is shown.  On the right plot, the address of the strip with the earliest arrival time outputted by the VMM1 compared to Monte Carlo simulation.}
\label{fig:trigg}
\end{figure}

In order to reduce the number of trigger channels the design takes advantage
of the micromegas detector's fine readout pitch to reduce the number of channels by
a factor of 64 resulting in a total channel count for the trigger logic of about 33,000. This is accomplished by considering only the first arriving hit in each 64-channel front-end ASIC for a given bunch crossing (25\,ns), resulting in a system with  granularity of 32\,mm (64\,$\times$\,0.5\,mm) but having spatial resolution better than half a millimetre by simply recording the address of the strip with the earliest arrival. This task is realized by a newly developed front-end ASIC, the VMM. Developed at the Brookhaven National Laboratory (BNL), the VMM1 (first version of the VMM family) was tested in 2012 test beam fulfilling the requirements of providing precision tracking and trigger data. Figure \ref{fig:trigg} shows the angular resolution achieved at the trigger level with six micromegas detector at 50\,cm apart and the strip with the earliest arrival time compared to Monte Carlo. The variation observed in the strip selection is due to ionizing fluctuation depending on the incident track angle.

\subsection{Ageing}\label{sec:ageingMM}
\begin{figure}[tbp] 
\centering
\includegraphics[width=.6\textwidth]{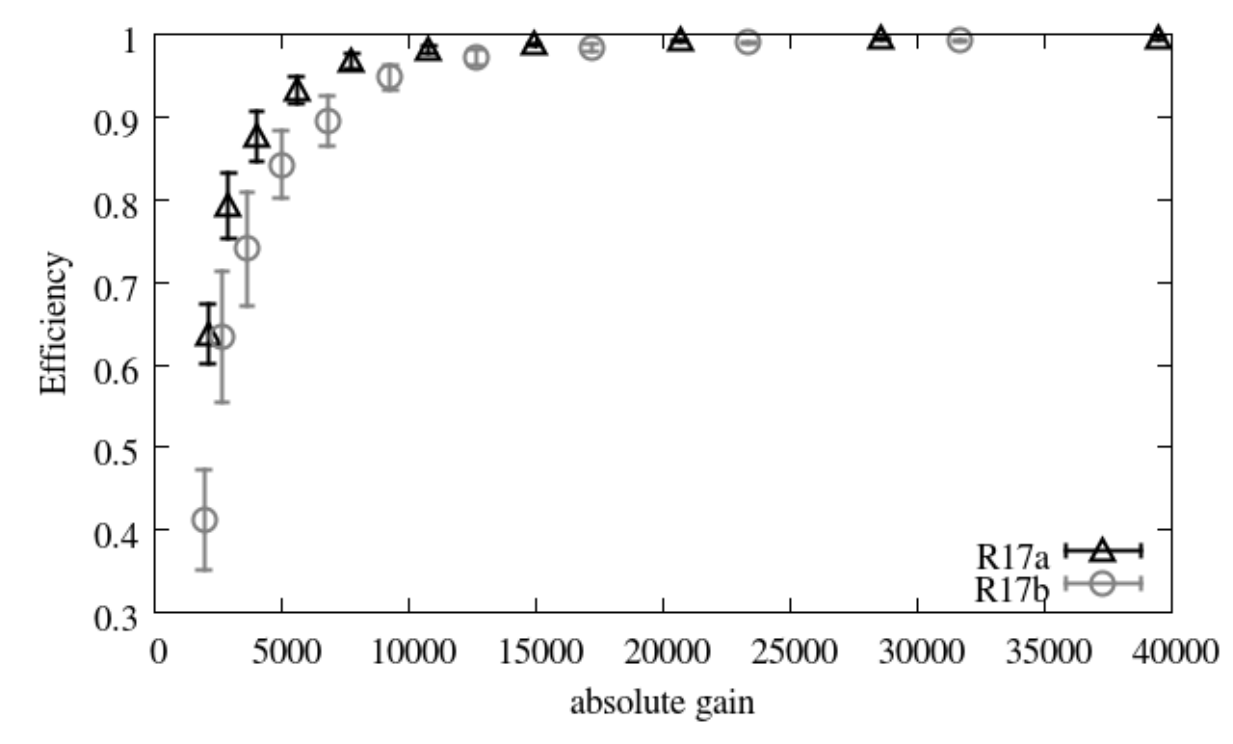}
\caption{The efficiency measurement as a function of the absolute gain for both irradiated and non-irradiated detectors.}
\label{fig:ageingGainCurves}
\end{figure}

Extensive ageing tests of a $10\times10\,\mathrm{cm^2}$ resistive-strip micromegas detector were performed at CEA Saclay in 2011 (ref.~\cite{cite:galanAgeing}) and 2012 with 8\,keV X-rays, thermal neutrons, $\sim$1\,MeV gamma rays and alpha particles. Eash of these exposures collected dose, shown in table~\ref{tab:radtests}, rates equivalent to the ones expected in the most exposed region of the NSW in about 10 years of LHC operation at a luminosity of $\mathcal{L}=5\times10^{35}\,\mathrm{cm^{-1}s^{-1}}$. A second detector, not irradiated, served as reference detector. No signs of performance deterioration of the exposed detector were observed. The efficiency measurement as a function of the absolute gain for both irradiated and non-irradiated detectors is shown in figure~\ref{fig:ageingGainCurves}.

\begin{table}[tbp]
\caption{Radiation test with micromegas detectors.}
\label{tab:radtests}
\smallskip
\centering
\begin{tabular}{| l | c | l |}
\hline
Irradiation with \raisebox{1.5ex}& Charge Deposit (mC/cm$^2$) & HL-LHC Equivalent\\
\hline\hline
X-Ray&225&5 HL--LHC years equivalent\\[1ex]
\hline
X-Neutron&0.5&10 HL--LHC years equivalent\\[1ex]
\hline
Gamma&14.84&10 HL--LHC years equivalent\\[1ex]
\hline
Alpha&2.4&$5\times10^8$ sparks equivalent\\[1ex]
\hline
\end{tabular}
\end{table}

\subsection{Testing Micromegas in ATLAS Cavern}\label{sec:atlasMM}
\begin{figure}[h]
	\begin{center}
		\includegraphics[scale=.51]{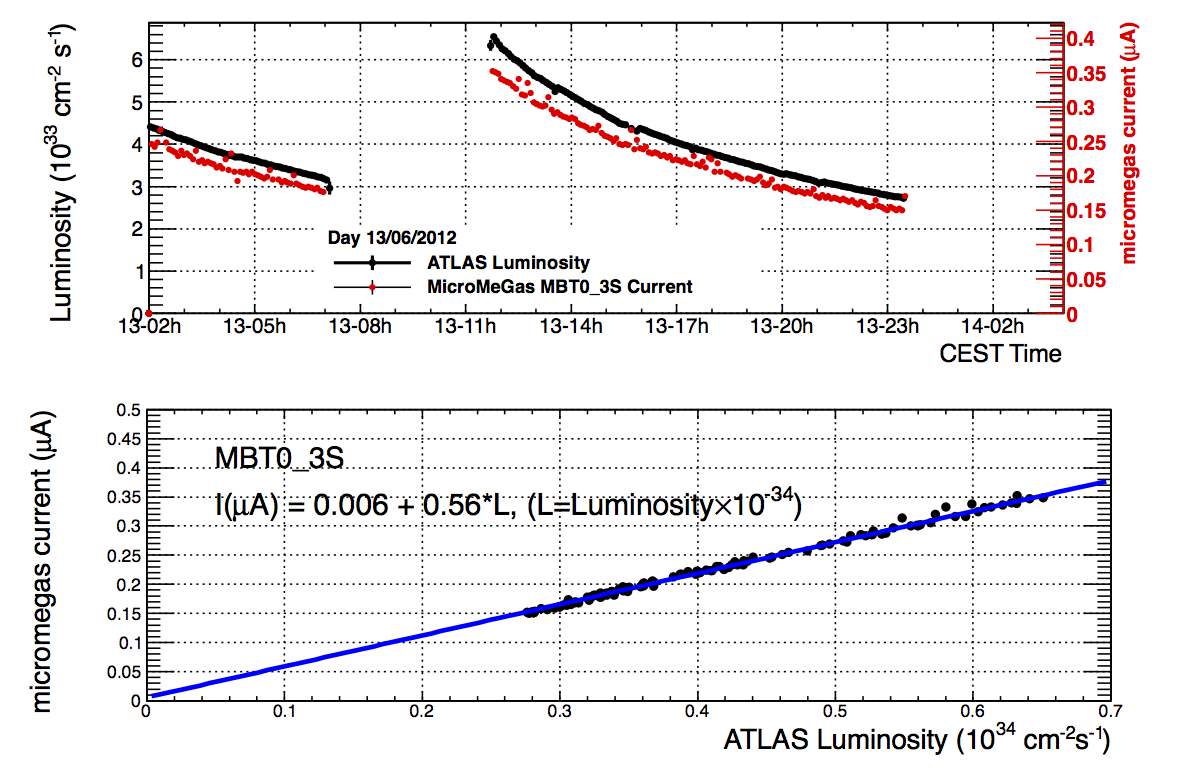}
		\caption{The top plot shows the MBT current (red points) and the ATLAS luminosity (black line). The lower plot gives the MBT current versus the ATLAS luminosity. The blue line is a linear fit to the data.}
		\label{fig:day7}
	\end{center}
\end{figure}

Five small prototype micromegas detectors were installed in the ATLAS detector during LHC running at $\sqrt{s} = 8\, \mathrm{TeV}$ realizing the operation of micromegas detectors in a real high energy experiment. One two-gap $9\times 4.5\, \mathrm{cm^2}$ area detector (MBT) was placed in front of the electromagnetic calorimeter, 1\,m radially and 3.5\,m horizontally from the interaction point. On the surface of the electromagnetic calorimeter particle rates at the level of  $70\,\mathrm{kHz/cm^2}$ at ATLAS luminosity $\mathcal{L}=5\times 10^{33}\,\mathrm{cm^{-2}s^{-1}}$ are expected.  Four $9\times 10\, \mathrm{cm^2}$ detectors installed on the current  ATLAS Small Wheels at the inner region in front of the Cathode Strip Chambers (CSC). A comparison of the currents drawn by the detector installed in front of the electromagnetic calorimeter with the luminosity measurement in ATLAS experiment was done showing a strong correlation between them. Figure \ref{fig:day7} shows the MBT current together with the ATLAS luminosity for one day of data taking. The four micromegas on the small wheel operated under much less particle flux reconstructing particle tracks without any problem.

\section{Conclusions}\label{sec:conclusions}
The extensive MAMMA R\&D program has transformed micromegas detectors spark resistant allowing them to operate in high energy experiments like ATLAS. Construction techniques developed, allowed us to build large area detectors of 2\,m$^2$ while involving industry. The first generation of the VMM ASIC unveiled the capabilities of micromegas as a trigger and tracking detector. All the above achievements brought micromegas detectors ahead of the R\&D phase. A full sector identical to a detector that can be installed to ATLAS will be build in 2014 featuring the next generation of the VMM ASIC providing trigger and tracking information. Following the ATLAS schedule, 1200\,m$^2$ of micromegas detectors will be constructed and assembled on 2015--2016 while the installation and commission of the full system will follow on 2017--2018.

\acknowledgments
The work described here is an intensive work performed from the MAMMA collaboration during the last years. The CERN PCB workshop team played a major role in development and construction of almost all micromegas prototypes and a big thanks goes to them as well.

The present work was co-funded by the European Union (European Social Fund ESF) and Greek national funds through the Operational Program "Education and Lifelong Learning" of the National Strategic Reference Framework (NSRF) 2007-1013. ARISTEIA-1893-ATLAS MICROMEGAS.




\end{document}